\documentclass[apjl]{emulateapj}

\usepackage{apjfonts}

\newcommand{\eg}{e.g., }
\newcommand{\ie}{i.e., }
\newcommand{\Msun}{M_{\odot}}
\newcommand{\kms}{km~s$^{-1}$}

\newcommand{\Nifs}{$^{56}$Ni}

\newcommand{\Mej}{M_{\rm ej}}
\newcommand{\KE}{E_{\rm K}}
\def\gsim{\mathrel{\rlap{\lower 4pt \hbox{\hskip 1pt $\sim$}}\raise 1pt
\hbox {$>$}}}
\def\lsim{\mathrel{\rlap{\lower 4pt \hbox{\hskip 1pt $\sim$}}\raise 1pt
\hbox {$<$}}}

\newcommand{\samurai}{\texttt{SAMURAI}}

\def\ion#1#2{{\rm #1}~{\sc #2}}

\shorttitle{Multi-Dimensional Simulations for Hypernova Spectra}
\shortauthors{Tanaka et al.}
 
\begin{document}

\title{
Multi-Dimensional Simulations for Early Phase Spectra of 
Aspherical Hypernovae: \\
SN 1998\lowercase{bw} and Off-Axis Hypernovae}
\author{
Masaomi Tanaka \altaffilmark{1},
Keiichi Maeda \altaffilmark{2,3},
Paolo A. Mazzali \altaffilmark{2,4,1}, and
Ken'ichi Nomoto \altaffilmark{1}
}

\altaffiltext{1}{Department of Astronomy, Graduate School of Science, University of Tokyo, Hongo 7-3-1, Bunkyo-ku, Tokyo 113-0033, Japan; mtanaka@astron.s.u-tokyo.ac.jp}
\altaffiltext{2}{Max-Planck-Institut f\"{u}r Astrophysik, Karl-Schwarzschild-Str. 1, D-85741 Garching bei M\"{u}nchen, Germany; maeda@MPA-Garching.MPG.DE}
\altaffiltext{3}{Department of Earth Science and Astronomy,Graduate School of Arts and Science, University of Tokyo, Meguro-ku, Tokyo 153-8902, Japan}
\altaffiltext{4}{Istituto Nazionale di Astrofisica, OATs, Via Tiepolo 11, I-34131 Trieste, Italy}

\begin{abstract}

Early phase optical spectra of aspherical jet-like supernovae (SNe) 
are presented. 
We focus on energetic core-collapse SNe, or hypernovae.
Based on hydrodynamic and nucleosynthetic models,
radiative transfer in SN atmosphere is solved 
with a multi-dimensional Monte-Carlo radiative transfer code, \samurai.
Since the luminosity is boosted in the jet direction, the temperature 
there is higher than in the equatorial plane by $\sim$ 2,000 K.
This causes anisotropic ionization in the ejecta.
Emergent spectra are different depending on viewing angle, reflecting
both aspherical abundance distribution and anisotropic ionization. 
Spectra computed with an aspherical explosion model with kinetic energy
$20 \times 10^{51}$ ergs are compatible with those of the Type Ic SN 1998bw if 
$\sim 10-20$\% of the synthesized metals are mixed out to higher velocities.
The simulations enable us to predict the properties of off-axis hypernovae.
Even if an aspherical hypernova explosion is observed from the side, it should 
show hypernova-like spectra but with some differences in the line velocity, 
the width of the Fe absorptions and the strength of the \ion{Na}{i} line.

\end{abstract}

\keywords{supernovae: general --- supernovae: individual (SN~1998bw) 
--- radiative transfer}

\section{Introduction}
\label{sec:introduction}

The connection between the long gamma-ray bursts (GRBs) and 
a special class of Type Ic supernovae (GRB-SNe) is now well established
(Galama et al. 1998; Hjorth et al. 2003; Stanek et al. 2003;
Malesani et al. 2004; Pian et al. 2006).
Since GRBs are induced by relativistic jets, 
GRB-SNe are also expected to be aspherical.
There is also increasing evidence that core-collapse SNe are not spherically 
symmetric, 
coming from, \eg the detection of polarization in several SNe
(\eg Wang et al. 2001; Kawabata et al. 2002; Leonard et al. 2006)
and late-time spectroscopy (Mazzali et al. 2005).

However, the progenitors, the explosion mechanisms
and the origins of diverse properties of GRB-SNe 
are still not understood (\eg Nomoto et al. 2007).
In order to answer such questions, it is crucial that the properties of the
explosions (\eg the mass and kinetic energy of the ejecta and the asymmetry 
of the explosion) are accurately derived from observations.
For this purpose, we should know the observational properties of 
multi-dimensional SN explosions,
and to what degree they are affected by orientation effects.

This is an area that has not been studied in great depth, despite its importance.
Although some works have addressed the properties of 
the light curve (LC) or the late phase spectra of aspherical SNe
(\eg H\"oflich et al. 1999; Maeda et al. 2002; 
Maeda et al. 2006a, c; Sim 2006), only a few studies of early phase spectra 
($t \lsim 50$ days, where $t$ is the time since the explosion) have been 
performed (\eg H\"oflich et al. 1996 for core-collapse SNe, 
Thomas et al. 2002; Kasen et al. 2004; Tanaka et al. 2006 for Type Ia SNe).

SNe that associate with GRBs are thought to be highly energetic 
explosions, \ie hypernovae (here defined as SNe with ejecta kinetic energy 
$E_{51} = \KE / 10^{51} {\rm ergs} > 10$; \eg Nomoto et al. 2006)
indicated by broad line features in their early phase spectra.
However, this suggestion was based on analyses 
that assumed spherical symmetry. 
No realistic multi-dimensional explosion models have been 
verified against the observed early phase spectra.

We report the first study of early phase spectra of 
aspherical hypernova models using a multi-dimensional radiative transfer code.
The synthetic spectra are compared with observed spectra of SN 1998bw, and
implications for off-axis hypernovae are discussed.

\section{Models}
\label{sec:models}

We use the results of multi-dimensional hydrodynamic 
and nucleosynthetic calculations for SN 1998bw (Maeda et al. 2002) 
as input density and element distributions.
In the present simulations, 16 elements are included,
\ie H, He, C, N, O, Na, Mg, Si, S, Ti, Cr, Ca, Ti, Fe, Co and Ni.
Since the original models used a He star as a progenitor,
we simply replace the abundance of the He layer with that of the C+O layer.
In the hydrodynamic model, energy is deposited aspherically, with 
more energy in the jet direction (z-axis).
As a result, \Nifs\ is preferentially synthesized along this direction 
(see Fig. 3 of Maeda et al. 2002).
In this {\it Letter}, an aspherical model with $E_{51} =20$ (A20)
and two spherical models with $E_{51} =20$ and $50$ 
(F20 and F50, respectively) are studied (Table \ref{tab:model}).
They are constructed based on the models with $E_{51}=10$ 
(Maeda et al. 2002; Maeda et al. 2006c).

Bolometric light curves (LCs) computed in 3D space 
by Maeda et al. (2006c) 
are also used as input for the spectral calculations.
A common problem in modeling hypernova LCs is  
that a spherical model reproducing the early rise of the LC 
($E_{51} = 50$ for SN 1998bw) declines more rapidly than the observed LC
at $t \sim 100 - 150$ days (Nakamura et al. 2001; Maeda et al. 2003).
This problem can be solved by aspherical models with a polar view.
In aspherical models, even with a lower kinetic energy
($E_{51} = 10 - 20$) than in the spherical case ($E_{51} = 50$),
which allows sufficient trapping of $\gamma$-rays at late times,
the rapid rise of the LC can be reproduced
because of the extended \Nifs\ distribution (Maeda et al. 2006c).

\begin{deluxetable}{lcccc} 
\tablewidth{0pt}
\tablecaption{Summary of Explosion Models}
\tablehead{
model name & 
$\Mej$ \tablenotemark{a} & 
$E_{51}$  &
$M($\Nifs$)$ \tablenotemark{b} &
$p$ \tablenotemark{c}
}
\startdata
A20          & 10 & 20 & 0.39 & 0.0  \\
F20          & 10 & 20 & 0.31 & 0.0  \\
F50          & 10 & 50 & 0.40 & 0.0  \\ \hline
A20p0.2      & 10 & 20 & 0.39 & 0.2  \\ 
\enddata
\tablenotetext{a}{The mass of the ejecta ($\Msun$)} 
\tablenotetext{b}{The ejected \Nifs\ mass ($\Msun$)}
\tablenotetext{c}{The mixing fraction (see \S \ref{sec:comparison})}
\label{tab:model}
\end{deluxetable}

\section{Method}
\label{sec:method}

In order to study the detailed properties of the radiation 
from aspherical SNe, 
we have unified a SupernovA MUlti-dimensional RAdIative transfer code \samurai.
\samurai\ is a combination of 3D codes adopting Monte-Carlo 
methods to compute the bolometric LC (Maeda et al. 2006c), and
the spectra of SNe from early (Tanaka et al. 2006) to late phases 
(Maeda et al. 2006a).
The LCs are computed as pioneered by Cappellaro et al. (1997).
Our 3D LC code adopts individual packet method as described by Lucy (2005) 
and additionally, 
includes multi-energy transfer for high energy photons (Maeda 2006b).

The early phase spectra are calculated as snapshots 
in the optically thin atmosphere, using the results of 
the LC simulation as initial conditions.
A sharply defined photosphere is assumed as an inner boundary for simplicity.
The position of the inner boundary in each direction is 
determined by averaging the positions of the last scattering photon packets 
(see Fig. 3 of Maeda et al. 2006c).
Thus, the inner boundary corresponds to an optical depth of roughly unity. 
The luminosity at the inner boundary ($L_{\rm in} (\theta)$) 
is also taken from the LC computation by integrating the energy of
photon packets escaping from each $\theta$ bin.

For the computation of ionization and excitation state
in the atmosphere,
the local physical process same as in the previous 1D code 
(Mazzali \& Lucy 1993; Mazzali 2000), which is used to model 
spectra of SNe of various types (\eg Mazzali et al. 1993), is adopted.
Line scattering under the Sobolev approximation
and electron scattering are taken into account.
For line scattering, the effect of photon branching is included
as in Lucy (1999).
Beginning with a trial temperature structure, 
a number of photon packets (typically $\sim 10^8 - 10^9$) 
are followed in 3D space, giving the flux in each mesh.
Then the temperature structure is updated.
After the temperature structure converges, 
the spectra are obtained by counting the energy (or wavelength)
distribution of escaping photon packets in every direction.

Since the background model is axially symmetric, it is not 
necessary to follow the propagation in the azimuthal direction explicitly.
We use a 2D grid ($r,\ \theta$), 
but still perform the transport of the photon packets in 3D,
reproducing photon propagation in the azimuthal direction 
by projecting the photon back on the 2D plane of the calculation.
The spatial grid has a typical mesh number $(n_r, n_{\theta}) = (20, 80)$.
Velocity is used as a proxy for the radial coordinate thanks  
to the homologous expansion of SN ejecta ($r = vt$).
Details of the code will be presented elsewhere.

\section{Results}
\label{sec:results}

\begin{figure}
\begin{center}
\includegraphics[scale=0.5]{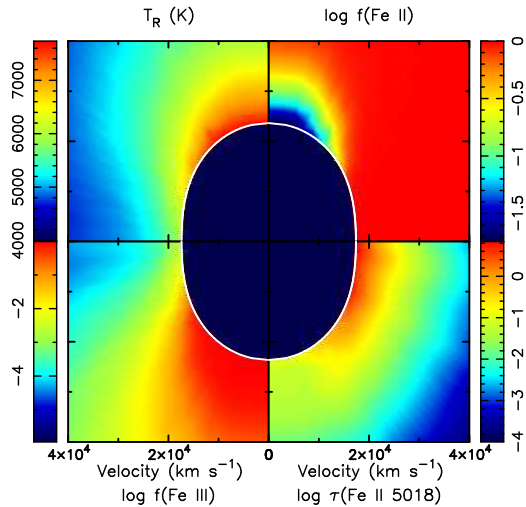}
\caption{
{\it Upper left}: Computed temperature structure in the atmosphere
of model A20 at 20 days after the explosion.
{\it Upper right, lower left}: Ionization fractions of 
\ion{Fe}{ii} and \ion{Fe}{iii}.
{\it Lower right}: Sobolev optical depth of the \ion{Fe}{ii} $\lambda$5018.
The vertical axis (z-axis) is the jet direction.
\label{fig:tempiond20}}
\end{center}
\end{figure}

At $t=20$ days, 
the photosphere of model A20 is mildly aspherical with an axis ratio of 
1.4 between the polar and equatorial directions.
The emergent luminosity is also anisotropic, being brighter by a factor of 
1.2 in the polar direction (Maeda et al. 2006c).
In contrast, the local luminosity at the photosphere, $L_{\rm in} (\theta)$, 
is highly aspherical, with an axis ratio of 6.3, tracing the aspherical 
\Nifs\ distribution.
At this epoch, the photosphere is still outside the region where explosive 
nucleosynthesis occurs.
The metal lines in the spectrum result from pre-explosion abundance in the C+O 
layer.

The upper left panel of Figure \ref{fig:tempiond20} shows the calculated 
temperature structure.
The boosted luminosity makes the temperature near the z-axis higher than in 
the equatorial plane by $\sim 2000$ K.
This anisotropy of the temperature causes anisotropies of the ionization
structure.
The upper right and lower left panels of Figure \ref{fig:tempiond20} show 
the ionization fractions of \ion{Fe}{ii} and \ion{Fe}{iii}, respectively.
Because of the high temperature, \ion{Fe}{ii} is suppressed 
by a factor of more than 30 near the z-axis. 
As a result, the optical depth of the \ion{Fe}{ii} lines is larger in the 
equatorial plane (Fig. \ref{fig:tempiond20}, lower right panel).

\begin{figure}
\begin{center}
\includegraphics[scale=1.2]{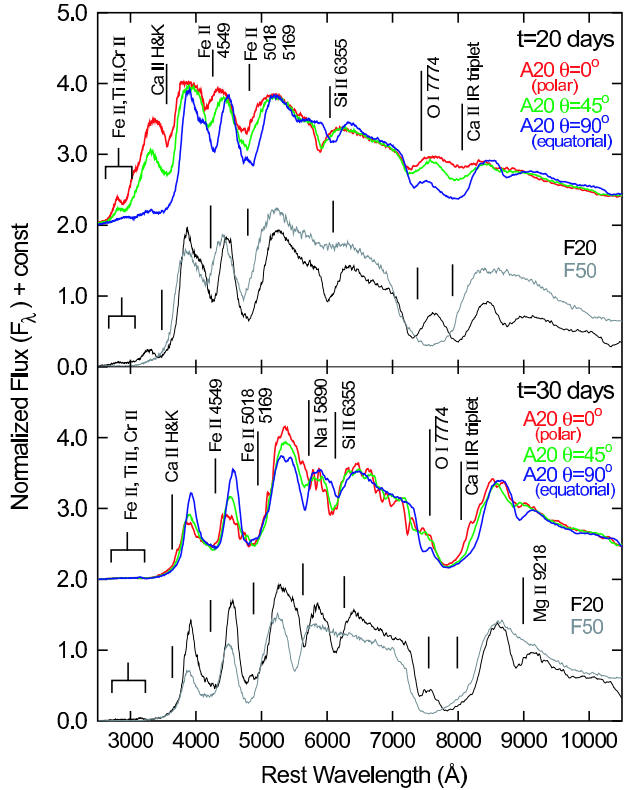}
\caption{
{\it Upper}: Synthetic spectra of model A20, F20 (black) and F50 (gray)
at 20 days after the explosion.
The red, green and blue lines show the spectra of model A20 seen 
from $\theta = 0^{\circ}, 45^{\circ}$ and $90^{\circ}$, respectively.
Here $\theta$ is measured from the z-axis (jet direction).
{\it Lower}: Same as upper panel but at 30 days after the explosion.
\label{fig:specall}}
\end{center}
\end{figure}

The upper panel of Figure \ref{fig:specall} shows the synthetic spectra
at $t=20$ days for models A20 (red, green and blue 
for $\theta = 0^{\circ}$ (polar), $45^{\circ}$ and $90^{\circ}$ 
(equatorial), respectively), F20 (black) and F50 (gray).
Here $\theta$ is measured from the z-axis.
All absorption lines except for \ion{Si}{ii} $\lambda$6355 are
stronger for larger $\theta$, \ie for an equatorial view.
This is because all species that have strong lines,
\ie \ion{O}{i}, \ion{Si}{ii}, \ion{Ca}{ii}, \ion{Ti}{ii}, \ion{Cr}{ii}
and \ion{Fe}{ii}, dominate near the equator but not near the z-axis
as shown in Figure \ref{fig:tempiond20} for the case of Fe.
\footnote{\ion{Si}{ii} $\lambda$6355 is not strong 
at $\theta = 90^{\circ}$ because the temperature 
in the equatorial plane is too low to activate the transition.}. 
This also leads to a more effective flux-blocking in the near-UV 
in the side-viewed spectrum.

\begin{figure}
\begin{center}
\includegraphics[scale=0.5]{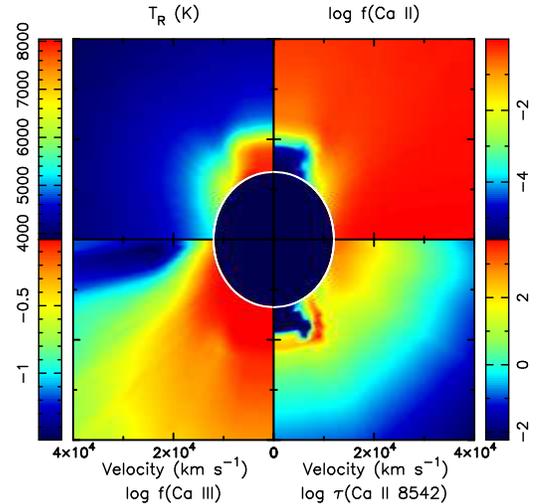}
\caption{
{\it Upper left}: Computed temperature structure in the atmosphere
of model A20 at 30 days after the explosion.
{\it Upper right, lower left}: Ionization fractions of 
\ion{Ca}{ii} and \ion{Ca}{iii}.
{\it Lower right}: Sobolev optical depth of \ion{Ca}{ii} 
$\lambda$8542.
\label{fig:tempiond30}}
\end{center}
\end{figure}

The lower panel of Figure \ref{fig:specall} shows the synthetic 
spectra at $t=30$ days.
At this epoch, the local luminosity and the temperature structure are still 
anisotropic, although the photosphere is almost spherical 
(upper left panel of Fig. \ref{fig:tempiond30}).
Consequently, the distribution of ionization fractions is also aspherical 
(see Fig. \ref{fig:tempiond30}, upper right and lower left panels, 
for the case of Ca).
However, the emergent spectra are not significantly different for different 
viewing angles (Fig. \ref{fig:specall}).
This is the effect of the aspherical abundance distribution.
The photosphere at this epoch is located inside the region 
where heavy elements are synthesized in the explosion.
Since nucleosynthesis occurs entirely near the polar direction in our model,
the suppression of important ions near the z-axis is compensated by the larger 
abundance of the heavy elements
(see Fig. \ref{fig:tempiond30}, lower right panel)
\footnote{
The \ion{Fe}{ii} absorptions are broader at $t=30$ days 
than at $t=20$ days because of more efficient absorption 
both near the photosphere 
($v \sim 12,000$ \kms, a lower velocity than at $t=20$ days)
and at high velocity ($v \sim 25,000$ \kms, 
where the temperature is lower and 
the \ion{Fe}{ii} fraction is higher than at $t=20$ days).
}.
Only the \ion{Na}{i} $\lambda$5890 line is stronger 
for larger $\theta$ because of the combined effect of the lower temperature 
that favors \ion{Na}{i} and the higher abundance of Na, 
which is predominantly synthesized before the explosion.

Given the strong anisotropy in $L_{\rm in} (\theta)$, 
the side-viewed spectrum of model A20 consists mainly of photons
escaping from the ejecta of $\theta \sim 60^{\circ}$.
Since physical quantities there 
(\eg isotropic mass and kinetic energy of the ejecta in that direction, 
and $L_{\rm in} (\theta)$) are similar to those of model F20,
the spectrum of Model F20 is similar to the polar-viewed spectrum of model A20.
Model F50 shows a very strong \ion{Na}{i} $\lambda$5890 line owing 
to the low temperature.

\section{Comparison with SN 1998\lowercase{bw}}
\label{sec:comparison}

\begin{figure}
\begin{center}
\includegraphics[scale=1.35]{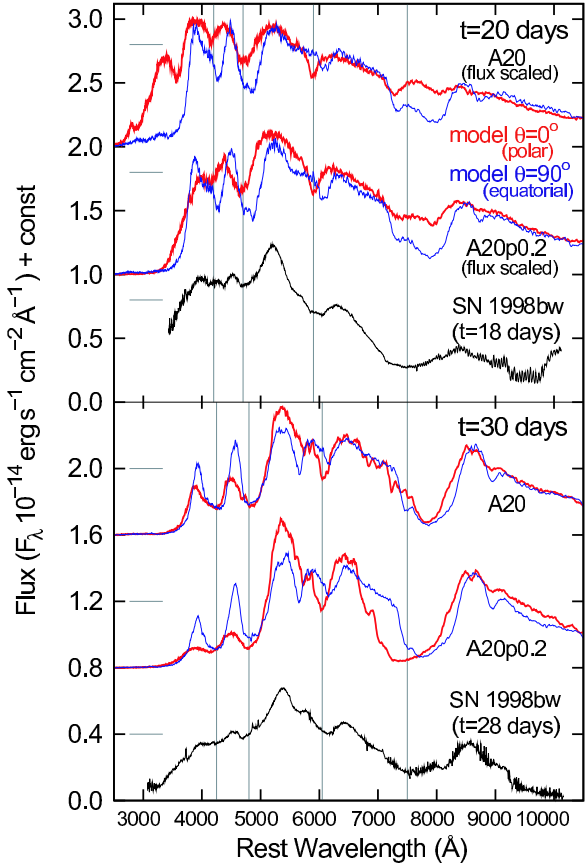}
\caption{
The spectra computed with model A20 and A20p0.2
compared with the spectra of SN 1998bw.
The synthetic spectra are reddened with $E(B-V)=0.032$ 
and scaled assuming $\mu = 32.76$.
{\it Upper}: Synthetic spectra at 20 days after the explosion 
(shifted upward by 2.0 and 1.0 $\times 10^{-14}$, respectively)
and the spectrum of SN 1998bw 18 days after GRB 980425.
The flux is reduced by a factor of 1.25 after the computation.
The vertical lines are drawn at 4200, 4700, 5900 and 7500 \AA.
The horizontal lines show the same flux level 
($8.0 \times 10^{-15}$).
{\it Lower}: Synthetic spectra at 30 days after the explosion 
(shifted upward by 1.6 and 0.8 $\times 10^{-14}$, respectively)
and SN 1998bw 28 days after GRB 980425.
The vertical lines are drawn at 4250, 4800, 6050 and 7500 \AA.
The horizontal lines show the same flux level  
($4.0 \times 10^{-15}$).
\label{fig:specp}}
\end{center}
\end{figure}

The spectra computed for the aspherical model A20 are compared with those 
of SN 1998bw in Figure \ref{fig:specp}.
Because of the detection of GRB 980425, it is tempting to compare the 
polar-viewed spectrum (red lines) with that of SN 1998bw.
Since the computed LC is brighter than the observation by a factor of 1.25 
at $t = 20$ days (Maeda et al. 2006c), the synthetic spectra at $t = 20$ days 
are scaled {\it after} the computation by this factor 
in Figure \ref{fig:specp}.

At $t=20$ days (Fig. \ref{fig:specp}, upper panel), 
the scaled synthetic spectrum of 
model A20 with $\theta = 0^{\circ}$ does not show the broad absorption trough 
between 4000 and 5000\AA\ that is typical of hypernovae such as SN 1998bw. 
This is because in the model Fe-line absorption is not effective at high 
velocities, making it possible for photons to escape effectively near 4500\AA. 
Similarly, the \ion{O}{i}-\ion{Ca}{ii} absorption at 7000 -- 8000 \AA\ 
is very weak in Model A20.
The Si line velocity in the model is, however, comparable to that in SN 1998bw.

At $t=30$ days (Fig. \ref{fig:specp}, lower panel), the \ion{Ca}{ii} and 
\ion{Fe}{ii} lines are somewhat stronger, although the 
\ion{O}{i}-\ion{Ca}{ii} absorption at 7000 -- 8000 \AA\ is still narrower than 
in the observed spectrum. 
The peaks around 4000 and 4500 \AA\ are partially suppressed for a polar view
by the high velocity absorption by the extended Fe near the jet, 
while they are strong in the synthetic spectrum for the equatorial view. 
The suppression of the peaks is similarly seen in the spectrum of SN 1998bw.
The \ion{Na}{i} $\lambda$5890 line is very weak in SN 1998bw, 
in analogy with the spectrum with $\theta=0^{\circ}$.

The strengths of the \ion{Ca}{ii} and \ion{Fe}{ii} lines at $t=20$ days 
can be increased if heavy elements synthesized 
in the explosion are mixed to outer layers.
In SN explosions, Rayleigh - Taylor (R-T) instabilities are expected to occur
\footnote{In Type Ic SNe, R-T instabilities could develop at the 
(C+O)/(Si+Ni) interface (Kifonidis et al. 2000).},
which could deliver the newly synthesized elements to higher velocities.
We simulate this possibility by introducing a parameter, 
the mixing fraction $p$.
If the mass ejected inside a conical 
section of the ejecta is $M_{\rm ej} (\theta)$, and $M_i (\theta)$ is the mass
of a certain element contained in the conical section, 
we simulate mixing by taking a fraction ($p$) of the mass 
of a newly synthesized element and
distributing this mass homogeneously in the conical section. 
Thus, in the outer layers, the newly synthesized element will
have an abundance $X_i (r,\ \theta) = p M_i (\theta) / M_{\rm ej} (\theta)$
(+ pre-SN abundance). 
In the deeper region, the abundances are 
redistributed by a mixing down of the unburned material.

In Figure \ref{fig:specp}, synthetic spectra of models with $p=0.2$ 
(model A20p0.2) are shown.
At $t=20$ days (the synthetic spectra are similarly scaled, 
upper panel of Fig. \ref{fig:specp}), although the absorption 
line at 7000 - 8000 \AA\ is still weaker than the observation, the Fe lines are
strong, causing the broad absorption at $\lsim 5000$ \AA\ followed by the  
peak at $\sim$ 5200 \AA.
At $t=30$ days, the strong \ion{Ca}{ii} feature contributes to the broad 
absorption at 7000 - 8000 \AA, which is comparable to that in SN 1998bw.

Although the agreement of the spectra is far from perfect, 
the spectra of the model with mixing are in qualitative agreement 
with those of SN 1998bw.
Since we use multi-dimensional hydrodynamic and nucleosynthetic models, 
which are computed ab initio, there are not many parameters 
that we can freely control.
The agreement could be improved by additional modifications,
but they would be complicated and are beyond the scope of this work.
Note that our input luminosity is higher than that of 
SN 1998bw at $t=20$ days, making the ejecta temperature too high. 
If a luminosity comparable to that of SN 1998bw is used, 
neglecting the possible displacement of the photosphere, 
the \ion{O}{i}-\ion{Ca}{ii} feature becomes as strong as the observation.
Even in this case, however, mixing with $p \sim 0.1$ is still required. 

\section{Conclusions}
\label{sec:conclusions}

We have presented the first detailed simulations of the early phase 
optical spectra of realistic jet-like hypernova models.
The emergent spectrum is different for different viewing angles.
The spectral properties are determined by the combination of 
aspherical abundances and anisotropic ionization states.

Considering the complexity of the problem, 
the synthetic polar-viewed spectra are in reasonable agreement 
with those of SN 1998bw,
when $\sim 10 - 20$\% of the synthesized material is mixed out to 
higher velocities.
The kinetic energy of an aspherical model that reproduces, at least
qualitatively, the spectra of SN 1998bw ($E_{51} = 20$) can be smaller than 
that of a well-fitting spherical model ($E_{51} = 50$).
This is consistent with previous results obtained from models of the LC and the
late time spectra (Maeda et al. 2006a \& c).

The simulations enable us to predict the 
properties of the early phase spectra of hypernovae viewed off-axis.
Compared with the spectra seen from the polar direction (red),
the spectra seen from the equatorial direction (blue) have 
(1) a slightly lower absorption velocity, 
(2) stronger peaks around 4000 and 4500 \AA\ and
(3) a stronger \ion{Na}{i} $\lambda$5890 line than in SN 1998bw.
SNe similar to SN 1998bw in ejecta mass, kinetic energy and \Nifs\ mass should 
always show spectral features of ``hypernovae'' or ``broad line supernovae'', 
but with some differences as described above, depending on the viewing angle.

\acknowledgements

M.T. would like to thank Nozomu Tominaga and Toshikazu Shigeyama
for fruitful discussion and useful comments.
This research has been supported in part by 
the Grant-in-Aid for Scientific Research (18104003, 18540231)
and the 21st Century COE Program (QUEST) from the JSPS and MEXT of Japan.
M.T. is supported through the JSPS Research Fellowships for Young Scientists.

\end{document}